\documentclass{myepl}

\usepackage{amsmath,amssymb,amsthm} 
 
\DeclareMathOperator{\e}{e}             % natural number
\DeclareMathOperator{\re}{Re}           % real part
\DeclareMathSymbol{\geqsymb}{\mathalpha}{AMSa}{"3E}
\def\geqs{\mathrel\geqsymb}
\newtheorem{theorem}{Theorem}
\newcommand{\field}[1]{\mathbb{#1}}
\newcommand{\Z}{\field{Z}\,}    % integers
\newcommand{\fP}{\field{P}}     % probability
\newcommand{\E}{\field{E}}      % expectation

\newcommand{\set}[1]{\{#1\}}                              % brackets
         
\newcommand{\Bigset}[1]{\Bigl\{#1\Bigr\}}         
\newcommand{\biggset}[1]{\biggl\{#1\biggr\}} 
\newcommand{\brak}[1]{[#1]}                              % square brackets
\newcommand{\bigbrak}[1]{\bigl[#1\bigr]}         
         
\newcommand{\biggbrak}[1]{\biggl[#1\biggr]} 
\newcommand{\bigpar}[1]{\bigl(#1\bigr)}                   % paranthesis
\newcommand{\biggpar}[1]{\biggl(#1\biggr)}        
 
\newcommand{\biggabs}[1]{\biggl|#1\biggr|} 
\newcommand{\Order}[1]{{\mathcal O}(#1)}                  % Order relation

\newcommand{\biggOrder}[1]{{\mathcal O}\biggl(#1\biggr)} 
\newcommand{\bigprob}[1]{\fP\bigl\{#1\bigr\}}

\newcommand{\expec}[1]{\E\{#1\}}                          % expectation

\newcommand{\xper}{x^{\ab{per}}} 
\newcommand{\TK}{T_{\ab{K}}} 
\newcommand{\ftrans}{f_{\ab{trans}}} 
\newcommand{\tftrans}{\tilde f_{\ab{trans}}} 
\newcommand{\prt}{p_{\ab{rt}}} 
\newcommand{\icx}{\ab{i}} 				% imaginary number

\title{Universality of residence-time distributions\\  
in non-adiabatic stochastic resonance}
\shorttitle{Residence times for stochastic resonance}
\author{Nils Berglund\inst{1}\thanks{E-mail: \email{berglund@cpt.univ-mrs.fr}} 
\and Barbara Gentz\inst{2}\thanks{E-mail: \email{gentz@wias-berlin.de}}}
\shortauthor{N. Berglund \and B. Gentz}
\institute{
  \inst{1} Centre de Physique Th\'eorique \thanks{Unit\'e Mixte de Recherche 
  (UMR 6207) du CNRS et des Universit\'es d'Aix Marseille 1, Aix Marseille 2 
  et Sud Toulon--Var, affili\'ee \`a la FRUMAM.}, CNRS Luminy \\
           Case 907, 13288~Marseille Cedex 9, France 
\\
  \inst{2} Weierstrass Institute for Applied Analysis and Stochastics \\
           Mohrenstr.~39, 10117~Berlin, Germany 
}
				   
\begin{document}

%%%%%%%%%%%%%%%%%%%%%%%%%%%%%%%%%%%%%%%%%%%%%%%%%%%%%%%%%%%%%%%%%%%%%%%%%%%%%%

\maketitle

\begin{abstract}
We present mathematically rigorous expressions for the  residence-time and
first-passage-time  distributions of a periodically forced Brownian
particle in a bistable potential. For a broad range of forcing frequencies
and amplitudes, the distributions are close to periodically modulated
exponential ones. Remarkably, the periodic modulations are governed by
universal functions, depending on a single parameter related to the forcing
period. The behaviour of~the distributions and their moments is analysed,
in particular in the low- and high-frequency limits.  

\bigskip
\noindent{\small{\it Date.\/} August 13, 2004.}

\noindent{\small{\it Revised.\/} January 3, 2005.}

\end{abstract}

%%%%%%%%%%%%%%%%%%%%%%%%%%%%%%%%%%%%%%%%%%%%%%%%%%%%%%%%%%%%%%%%%%%%%%%%%%%%%%

The amplification by noise of a weak periodic signal acting on a
multistable system is known as  stochastic resonance (SR). A simple example
of a system showing SR is an overdamped Brownian particle in a symmetric
double-well potential, subjected to deterministic periodic forcing as well
as white noise. Despite of the amplitude of the forcing being too small to
enable the particle to switch from one potential well to the other, such
transitions can be made possible by the additive noise. For sufficiently
large noise intensity, depending on the forcing period, the transitions
between potential wells can become close to periodic. This mechanism was
originally proposed by Benzi \etal\/ and Nicolis and 
Nicolis~\cite{BSV,BPSV,Nicolis} in order to offer an explanation for the
close-to-periodic occurence of the major Ice Ages. Since then, it has been
observed in a large variety of physical and biological systems (for reviews
see, {\it e.g.}\/,~\cite{WM,MW,GHM,Wellens}). 

Although much progress has been made in the quantitative description of the
phenomenon of SR, many of its aspects are not yet fully understood.
Mathematically rigorous results have so far been limited to the regimes of
exponentially slow forcing~\cite{Freidlin1,ImkellerPavlyu02}, or moderately
slow forcing of close-to-threshold amplitude~\cite{BG2,BG4}. 

One of the measures introduced in order to quantify SR is the
residence-time distribution, that is, the distribution of the random time
spans the Brownian particle spends in each potential well between
transitions. SR is characterized by the fact that residence times are more
likely to be close to odd multiples of half the forcing period  than not.
The residence-time distribution was first studied by Eckmann and Thomas for
a two-level system~\cite{ET}. For continuous systems, it has been
estimated, in the case of adiabatic forcing, by averaging the escape rate
for the frozen potential over the distribution of jump
phases~\cite{ZhouMossJung90,ChoiFoxJung98}. 

For larger forcing frequencies, however, the adiabatic approximation can no
longer be used. An alternative approach is to consider time as an
additional dynamic variable, which yields a two-dimensional problem. In the
absence of noise, the system has two stable periodic orbits, one
oscillating around each potential well, and one unstable periodic orbit,
which oscillates around the saddle and separates the basins of attraction
of the two stable orbits. The residence-time distribution is closely
related to the distribution of first passages of the stochastic process
through the unstable orbit. This problem was first investigated by Graham
and T\'el~\cite{GrahamTel84,GrahamTel85} and  Day~\cite{Day5,Day3,Day6},
and later by Maier and Stein~\cite{MS4} and others ({\it
e.g.}\/,~\cite{Lehman00,Dykman01}).

At first glance, however, this two-dimensional approach seems to produce a
paradoxical result. Indeed, it is known from the classical
Wentzell--Freidlin theory~\cite{VF69,VF70,FW} that the distribution of
first-passage locations through a periodic orbit looks uniform on the level
of exponential asymptotics~\cite{Day5}. This is due to the fact that
translations along the periodic orbit do not contribute to the cost in
terms of action functional. How can this fact be conciled with the
quasistatic picture, which yields residence times concentrated near odd
multiples of half the forcing period? Obviously, the answer has to lie in
the subexponential behaviour of the distribution of transitions. 

In this Letter, we extend previous results of~\cite{Day6,MS4,Lehman00} to a
mathematically rigorous expression for the first-passage-time distribution
up to multiplicative errors in the subexponential prefactor, valid for a
broad range of forcing periods~\cite{BG7,BG8}, from which we then deduce
the residence-time distribution.  A particularly interesting aspect of
the result is that both distributions are governed by universal periodic
functions, depending only on the period of the unstable periodic orbit times
its Lyapunov exponent. All the model-dependent properties of the
distribution can be eliminated by a deterministic time change. 

%%%%%%%%%%%%%%%%%%%%%%%%%%%%%%%%%%%%%%%%%%%%%%%%%%%%%%%%%%%%%%%%%%%%%%%%%%%%%%

\section{Assumptions}

We consider one-dimensional stochastic differential equations of the form 
\begin{equation}
\label{a1}
\upd x_t = -\frac{\partial}{\partial x} V(x_t,t) \upd t + \sigma \upd W_t, 
\end{equation}
where $W_t$ is a standard Wiener process, describing white noise, and the
small parameter $\sigma$ measures the noise intensity (the diffusion
constant being $D=\sigma^2/2$). The double-well potential $V(x,t)$ depends
periodically on time, with period $T$. The simplest example is 
\begin{equation}
\label{a2}
V(x,t) = \frac14 x^4 - \frac12 x^2 - A \sin(\omega t)x, 
\end{equation}
where the forcing has angular frequency $\omega=2\pi/T$ and amplitude
$|A|<\sqrt{4/27}$. 

Our results apply to a general class of $T$-periodic double-well potentials.
We assume that for each fixed $t$, $V(x,t)$ has two minima at
$X^{\ab{s}}_{1,2}(t)$ and a saddle at $X^{\ab{u}}(t)$, such that 
\begin{equation}
\label{a3}
X^{\ab{s}}_1(t) < c_1 < X^{\ab{u}}(t) < c_2 < X^{\ab{s}}_2(t)
\qquad \forall t
\end{equation}
for two constants $c_1, c_2$ (in the particular case of the
potential~\eqref{a2}, one can take $c_2=-c_1=1/\sqrt3$). Using Poincar\'e
maps, it is then straightforward to show that in the absence of
noise, the system~\eqref{a1} has exactly three periodic orbits, one of them
unstable and staying between $c_1$ and $c_2$, which we denote by 
$\xper(t)$. We introduce the notations 
\begin{equation}
\label{a4}
a(t) = -\frac{\partial^2}{\partial x^2} V(\xper(t),t)
\end{equation}
for the curvature of the potential at $\xper(t)$, and 
\begin{equation}
\label{a5}
\lambda = \frac1T \int_0^T a(t) \, \upd t
\end{equation}
for the Lyapunov exponent of the unstable orbit. We assume that $\lambda$ is
of order $1$, but $T$ can become comparable to Kramers' time. 

Finally, we need a non-degeneracy assumption for the system, which assures
that the action functional is minimized on a discrete set of paths, and
excludes symmetries other than time-periodicity~\cite{BG7}. In particular,
it should not be possible to transform the equation into an autonomous one
by a time-periodic change of variables. In the special case of the
potential~\eqref{a2}, this condition is met when $A\neq 0$. In addition, we
will assume that $|A|$ is of order $1$, while $\sigma^2 \ll |A|$. 

%%%%%%%%%%%%%%%%%%%%%%%%%%%%%%%%%%%%%%%%%%%%%%%%%%%%%%%%%%%%%%%%%%%%%%%%%%%%%%

\section{Results}

Assume the system starts at time $t_0$ in a given initial point in the
left-hand potential well. The residence-time distribution is closely
related to the distribution of the {\em first-passage time}\/, that is, the
(random) first time $\tau$ at which $x_t$ crosses the unstable periodic
orbit $\xper(t)$. 

Our main result states that the probability distribution of $\tau$ is
governed by the following function, in a sense made precise in Theorem~1
below. Let 
\begin{equation}
p(t,t_0) = \frac1N
Q_{\lambda T} \bigpar{{\theta(t)-|\ln\sigma|}}
\frac{\theta'(t)}{\lambda\TK(\sigma)}
\e^{-[\theta(t)-\theta(t_0)]/\lambda\TK(\sigma)}
\ftrans(t,t_0) 
\label{r1}
\end{equation}
where we use the following notations:
\begin{itemize}
\item	$\TK(\sigma)$ is the analogue of Kramers' time in the autonomous
case; it has the form  
\begin{equation}
\label{r1a}
\TK(\sigma) = \frac{C}{\sigma}\e^{\overline V/\sigma^2}, 
\end{equation}
where $\overline V$ is the constant value of the quasipotential on
$\xper(t)$. $\overline V$ can be computed by a variational method, as the
minimum of the action functional over all paths connecting the bottom of
the left-hand potential well to $\xper(t)$ (see~\cite{FW}). In the limit of
small forcing amplitude, $\overline V$ reduces to twice the potential
barrier height. The prefactor has order $\sigma^{-1}$ rather than
$1$~\cite{Lehman00b,Lehman00}, due
to the fact that most paths reach $\xper(t)$ through a bottleneck of width
$\sigma$ 
(the width would be larger if $|A|$ were not of order $1$~\cite{MS2}). 

\item	$Q_{\lambda T}(y)$ is the announced universal periodic function,
of period $\lambda T$; it has the explicit expression 
\begin{equation}
\label{r2}
Q_{\lambda T}(y) = 2\lambda T\sum_{k=-\infty}^\infty A(y - k\lambda T) 
\qquad
\tx{with}
\qquad
A(z) = \frac12 \e^{-2z} \exp\Bigset{-\frac12 \e^{-2z}}, 
\end{equation}
and thus consists of a superposition of identical asymmetric peaks,
shifted by a distance $\lambda T$. The average of $Q_{\lambda T}(y)$ over
one period is equal to $1$. 

\item	$\theta(t)$ contains the model-dependent part of the distribution;
it is an increasing function of $t$, satisfying
$\theta(t+T)=\theta(t)+\lambda T$, and is given by 
\begin{equation}
\label{r3}
\theta(t) = \tx{const} + \int_0^t a(s)\,\upd s - \frac12\ln\frac{v(t)}{v(0)}, 
\end{equation}
where $v(t)$ is the unique periodic solution of the differential equation
$\dot v(t) = 2 a(t) v(t) + 1$.  It is related to the variance of
Eq.~\eqref{a1} linearized around $\xper(t)$, and has the expression 
\begin{equation}
\label{r4}
v(t) = \frac1{\e^{2\lambda T}-1} \int_t^{t+T}
\exp\biggset{\int_s^{t+T}2a(u)\,\upd u}\, \upd s.
\end{equation}

\item 	$\ftrans(t,t_0)$ accounts for the initial transient behaviour
of the system; it is an increasing function satisfying 
\begin{equation}
\label{r5}
\ftrans(t,t_0) = 
\begin{cases}
\biggOrder{\exp\biggset{-\dfrac{L}{\sigma^2}
\dfrac{\e^{-\lambda (t-t_0)}}{1-\e^{-2\lambda (t-t_0)}}}}
& \tx{for $\lambda (t-t_0) < 2 |\ln\sigma|$} 
\vspace{2mm}
\\
%&\\
1 - \biggOrder{\dfrac{\e^{-\lambda (t-t_0)}}{\sigma^2}}
& \tx{for $\lambda (t-t_0) \geqs 2 |\ln\sigma|$} 
\end{cases}
\end{equation}
where $L$ is a constant, describing the rate at which the distribution in
the left-hand well approaches metastable equilibrium.  The transient term
thus behaves roughly like
$\exp\set{-L\e^{-[\theta(t)-\theta(t_0)]}/\sigma^2(1-\e^{-2[\theta(t)-\theta(t_0)]})}$.
However, $\ftrans(t,t_0)$ can be different when starting with an initial
distribution that is not concentrated in a single point.

\item	$N$ is the normalization, which we compute below.  
\end{itemize}

The precise formulation of our result is the following:

\begin{theorem}
For any initial time $t_0$, any $\Delta\geqs\sqrt\sigma$, and all times
$t\geqs t_0$, 
\begin{equation}
\label{r6}
\bigprob{\tau\in[t,t+\Delta]} = 
\int_t^{t+\Delta} p(s,t_0) \,\upd s \;[1+r(\sigma)], 
\end{equation}
where $r(\sigma) = \Order{\sqrt\sigma}$. 
\end{theorem}

If it were not for the limitation on $\Delta$, which is due to technical
reasons, this result would show that the probability density of $\tau$ is
given by $p(t,t_0) [1+r(\sigma)]$. We expect the remainder to be of order
$\sigma$ rather than $\sqrt\sigma$. 
This result has been derived in~\cite{BG7} in the simplified setting of a
piecewise quadratic potential, with explicit values for $\overline V$, $C$
and $L$, $r(\sigma)=\sigma$, and no restriction on $\Delta$.  A full proof
for the general case will be given in~\cite{BG8}. 

The main idea behind the proof is that sample paths reaching $\xper(t)$,
say, during a time interval $[t,t+\Delta]\subset[nT,(n+1)T]$, are
concentrated in a neighbourhood of $n$ deterministic paths, the most
probable exit paths, or MPEPs. Each of these paths contributes to the
probability~\eqref{r6}. The $k$th term of the sum~\eqref{r2} is the
contribution of a MPEP remaining inside the left-hand well for $n-k$
periods, and then idling along $\xper(t)$ during the remaining $k$ periods
(extending the sum from $k\in\set{0,\dots,n-1}$ to $\Z$ only results in an
error of order $\sigma$). The special form of  the sum~\eqref{r2},
involving double-exponentials, has been previously noted
in~\cite{Day3,MS4}  and~\cite{Lehman00}. 

%%%%%%%%%%%%%%%%%%%%%%%%%%%%%%%%%%%%%%%%%%%%%%%%%%%%%%%%%%%%%%%%%%%%%%%%%%%%%%

\section{Discussion}

Let us now analyse the expression~\eqref{r1} in more detail. 

Taking $\theta(t)/\lambda$ as new time variable eliminates the factor
$\theta'(t)/\lambda$ in the density. Thus $\theta(t)/\lambda$ can be
considered as a natural parametrization of time, in which one has to
measure the first-passage-time distribution in order to reveal its universal
character. We may thus henceforth assume that $\theta(t)=\lambda t$. 

The universal periodic function $Q_{\lambda T}$ depends only on the single
parameter $\lambda T$. For large $\lambda T$, it consists of well-separated
asymmetric peaks, while for decreasing $\lambda T$ these peaks overlap more
and more and $Q_{\lambda T}(y)$ becomes flatter. In fact, one can easily
compute the Fourier series of $Q_{\lambda T}$, which reads 
\begin{equation}
\label{d2}
Q_{\lambda T}(y) = \sum_{q\in\Z} 2^{\pi\icx q/\lambda T} 
\Gamma \biggpar{1+\frac{\pi\icx q}{\lambda T}} 
\e^{2\pi\icx qy/\lambda T}. 
\end{equation}
Since the Euler Gamma function $\Gamma$ decreases exponentially fast as a
function of the imaginary part of its argument,  $Q_{\lambda T}(y)$ is
close, for small $\lambda T$, to a sinuso\"\i d of mean value $1$ and
amplitude exponentially small in $1/2\lambda T$. 

\begin{figure}
\twoimages[height=4.0cm]{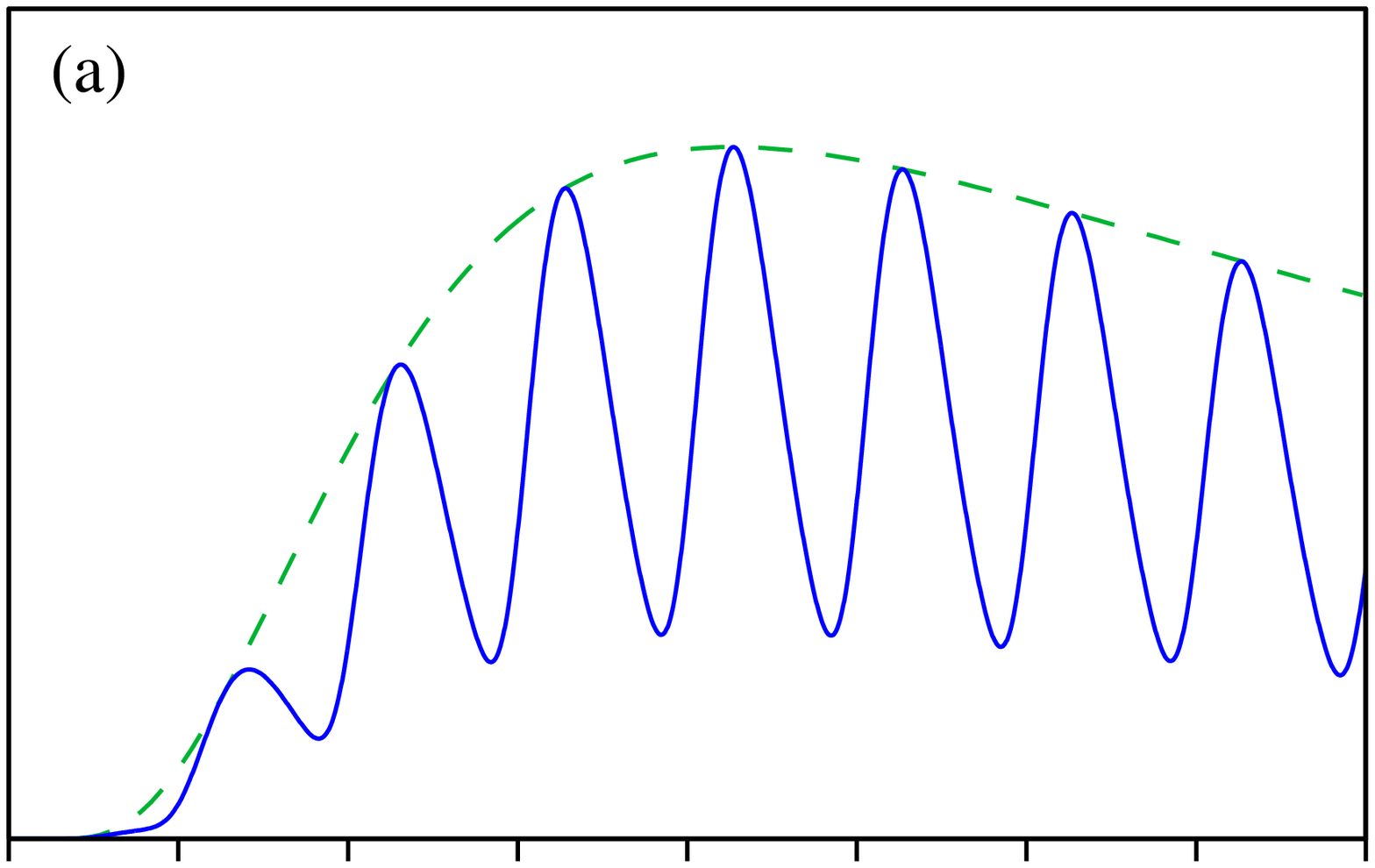}{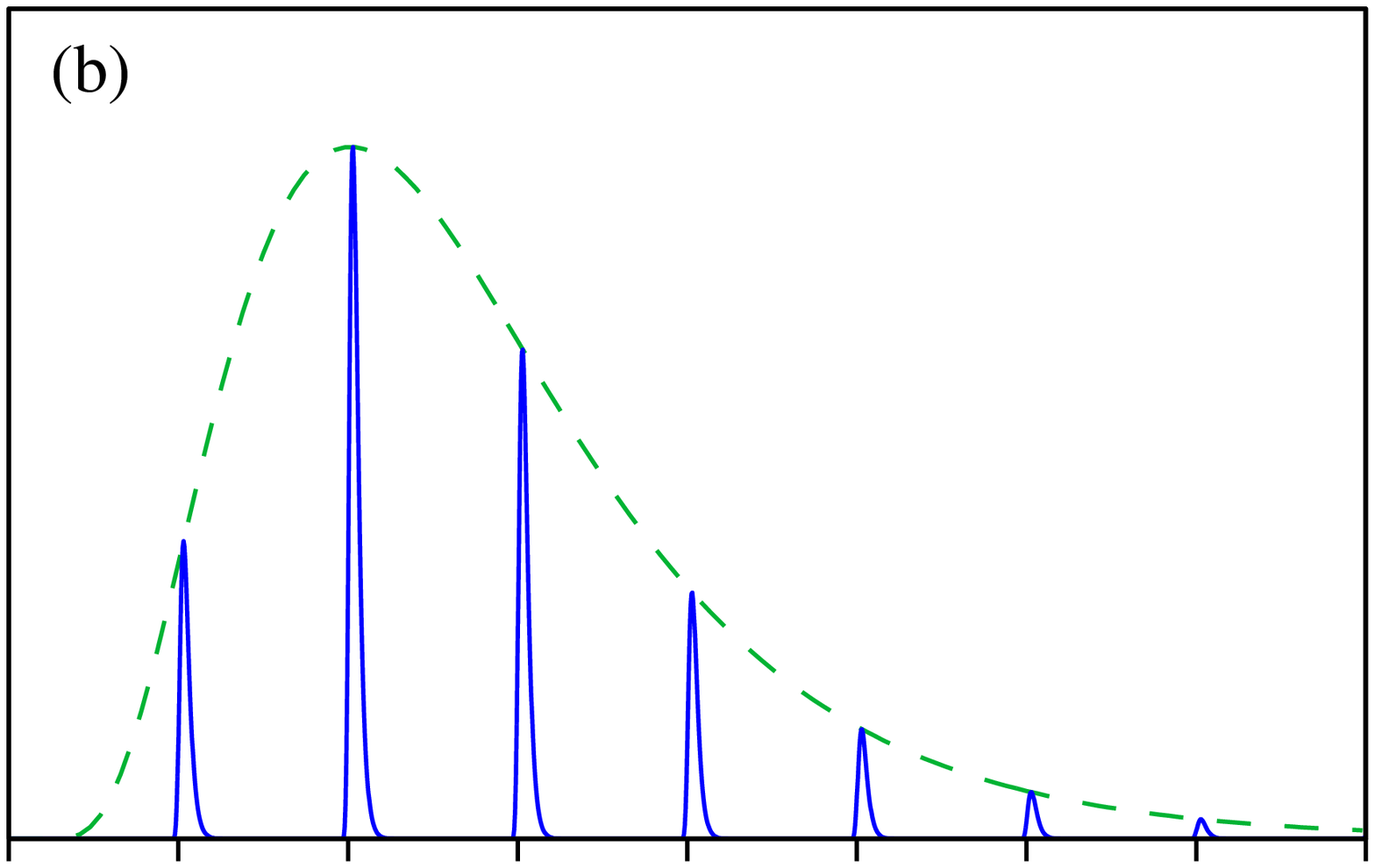}
\vspace{1mm}
\twoimages[height=4.0cm]{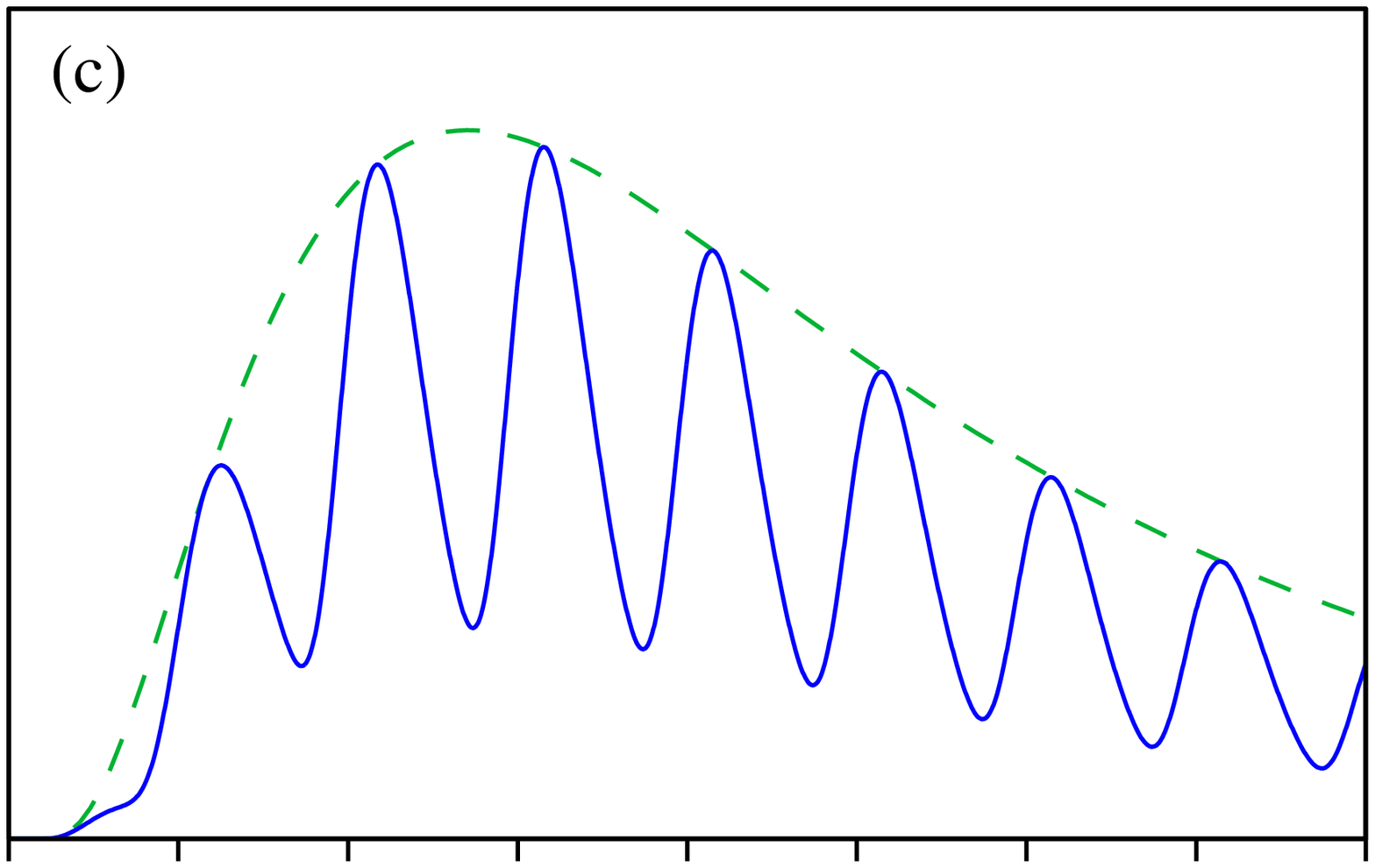}{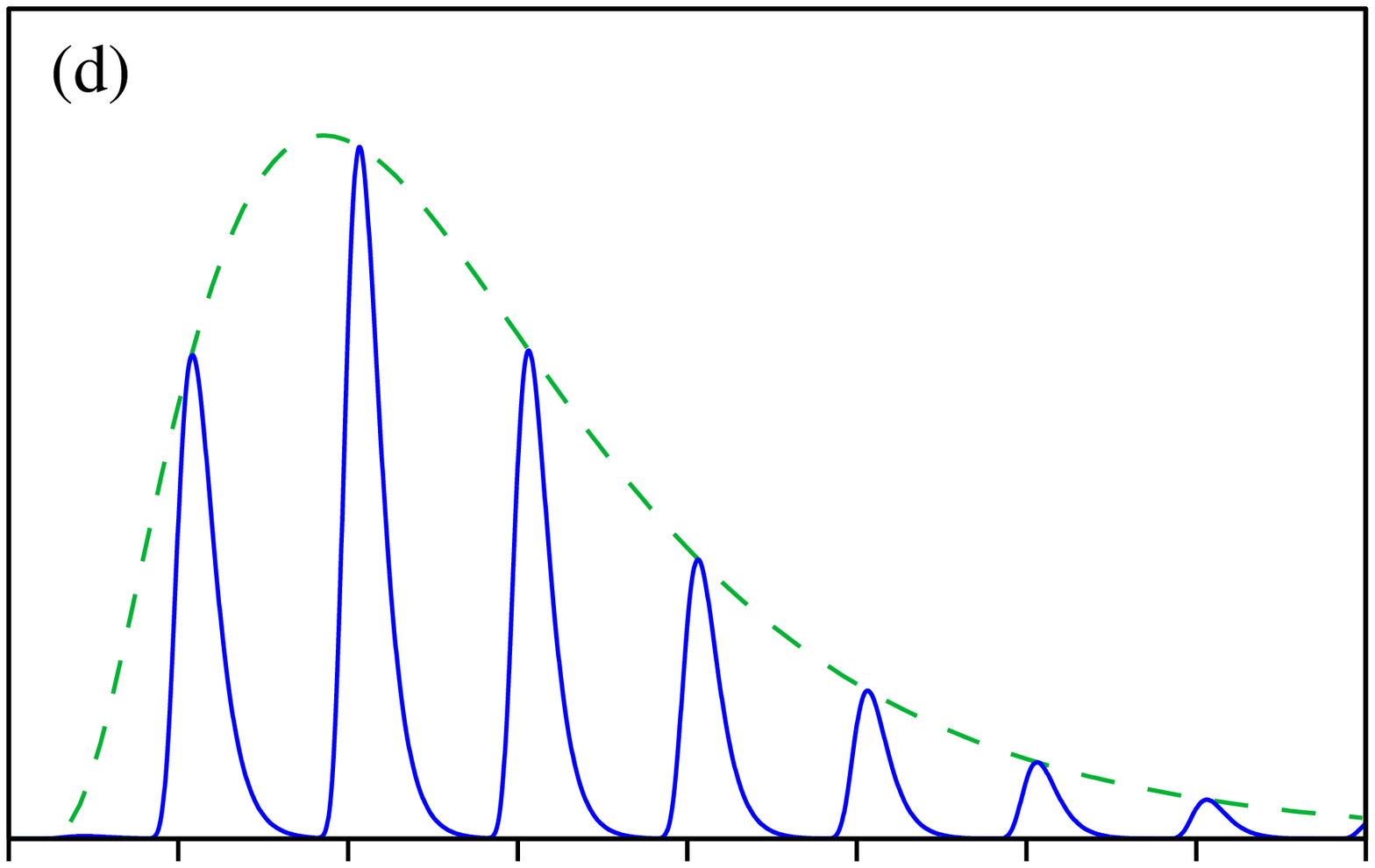}
\caption{
Plots of the first-passage-time distribution $p(t,0)$ (full curve) for
various parameter values. The broken curve is proportional to the average
density~\eqref{d4}, but scaled to match the peak height in order to guide
the eye. The $x$-axis comprises $8$ periods on each plot; the vertical
scale is not respected between plots. Parameter values are $\overline
V=0.5$, $\lambda=1$ and (a)~$\sigma=0.4$ ({\it i.e.}\/,
$D=\sigma^2/2=0.08$), $T=2$, (b)~$\sigma=0.4$, $T=20$, (c)~$\sigma=0.5$
({\it i.e.}\/, $D=0.125$), $T=2$ and (d)~$\sigma=0.5$, $T=5$.}
\label{fig1}
\end{figure}

The remarkable fact that $|\ln\sigma|$ enters in the argument of
$Q_{\lambda T}$ has been discovered, to our best knowledge, by Day, who
termed it {\em cycling}\/~\cite{Day3,Day6}. It means that as $\sigma$
decreases, the peaks of the first-passage-time distribution are translated
along the time-axis, proportionally to $|\ln\sigma|$. See also~\cite{MS4}
for an interpretation of this phenomenon in terms of MPEPs. 

The remaining, non-periodic time dependence of~\eqref{r1}  corresponds to
an averaged density, and behaves roughly like 
\begin{equation}
\label{d4}
\exp\biggset{-\frac{L}{\sigma^2} 
\frac{\e^{-\lambda (t-t_0)}}{1-\e^{-2\lambda (t-t_0)}} - \frac{t-t_0}{\TK(\sigma)}}. 
\end{equation}
This function grows from $0$ to almost $1$ in a time of order
$2|\ln\sigma|/\lambda$, and then slowly decays on the scale of the Kramers
time $\TK(\sigma)$. It is maximal for $\lambda (t-t_0) \simeq
\overline V/\sigma^2$. 

The first-passage-time distribution is thus in effect controlled by two
parameters: the quantity $\lambda T$, measuring the instability of the
saddle, which determines the shape of the distribution within a period; and
the Kramers time, which governs the decay of the average density~\eqref{d4}. 

For small $T/\TK(\sigma)$, the first-passage-time distribution consists of many
peaks whose height decreases only slowly on the time scale $\TK(\sigma)$.
Fig.\,\ref{fig1} shows first-passage-time distributions for relatively large
noise intensities, in order to make the decay more apparent. Increasing the
period for constant noise intensity has two effects (Fig.\,\ref{fig1}\,(a)
and~(b)): the peaks become narrower relatively to the period, while their
height decreases faster.  When $T$ increases beyond $\TK(\sigma)$, the
distribution becomes dominated by a single peak, and one enters the
synchronization regime, with the particle switching wells twice per period.
Increasing the noise intensity for constant period (Fig.\,\ref{fig1}\,(a)
and~(c)) also produces a faster decay of the peaks, while at the same time
the peak's location is shifted due to the cycling phenomenon.
Fig.\,\ref{fig1}\,(b) and~(d) correspond to the same value of
$T/\TK(\sigma)$, but in (d) a larger noise intensity is responsible for
broader peaks. 

Moments of the first-passage-time distribution can easily be computed up to a
correction stemming from $r(\sigma)$ (the correction due to $\ftrans(t)$ is
of smaller order). Using the Fourier series~\eqref{d2},
one finds (for $\theta(t)=\lambda t$) 
\begin{equation}
\expec{\tau^n} = \frac1N n!\, \TK(\sigma)^n  
\biggbrak{1+2\re \sum_{q\geqs1} 
\frac{(2\sigma^2)^{\pi\icx q/\lambda T}}
{(1-2\pi\icx q\TK(\sigma)/T)^{n+1}} 
\Gamma\biggpar{1+\frac{\pi\icx q}{\lambda T}}} 
\bigbrak{1+\Order{r(\sigma)}}.
\label{d5}
\end{equation}
In particular, taking $n=0$ yields the normalization $N$.  Note 
that  $\lim_{\sigma\to 0} \sigma^2\log\expec{\tau} = \overline V$, in
accordance with the classical Wentzell--Freidlin theory~\cite{VF69,VF70,FW}. 

Two other limits are of particular interest. For $\lambda T\ll 1$, the
decay properties of $\Gamma(1+\icx x)$ imply  $\expec{\tau^n} = n!\,
\TK(\sigma)^n \brak{1 + \Order{\e^{-\pi^2/2\lambda T}} +
\Order{r(\sigma)}}$,    which is close to the moments of an exponential
distribution with expectation $\TK(\sigma)$. This is natural since the
periodic modulation becomes flat in this limit. However, it is also true
that for $T\ll\TK(\sigma)$, one has $\expec{\tau^n} = n!\, \TK(\sigma)^n 
\brak{1 + \Order{T/\TK(\sigma)} + \Order{r(\sigma)}}$,  independently of
the value of $\lambda T$. This is due to the fact that the period of
modulation is short with respect to the scale of exponential decay. The
moments of the first-passage-time distribution can thus differ
significantly from those of an exponential distribution only when $T$ is
not too small compared to both $\lambda^{-1}$ and $\TK(\sigma)$. 

A third limit in which the first-passage-time distribution should approach an
exponential one is the limit of vanishing forcing amplitude $A$. However, the
expression~\eqref{r1} does not hold in cases where $A$ is not large
compared to $\sigma^2$, because it makes use of the saddle-point method in
the vicinity of MPEPs. An asymptotic expression for $A\ll\sigma^2$ has been
proposed in~\cite{ZhouMossJung90,ChoiFoxJung98}. 

Finally, the {\em residence-time distribution}\/ of the system can be
deduced from the knowledge of the first-passage-time distribution. Assume
that the Brownian particle makes a transition from the right-hand to the
left-hand potential well, crossing the unstable orbit at time~$s$, and,
having visited the left-hand well, crosses the unstable orbit again at
time~$s+t$. The residence-time distribution $\prt(t)$ is obtained~\cite{ET} by
integrating $p(s+t,s) \psi(s)$ over one period, 
where $\psi(s)$ is the asymptotic distribution of arrival {\em phases}\/ (i.e., times
modulo $T$). Assuming that, as for the potential~\eqref{a2}, the two wells
move half a period out of phase, 
$\psi(s) = Q_{\lambda T}(\lambda(s-T/2)-|\ln\sigma|)/T$. 
The Fourier series~\eqref{d2} yields, for $\theta(t)=\lambda t$ and up to a
multiplicative error $1+\Order{r(\sigma)}$, the residence-time distribution
\begin{equation}
\label{d9}
\prt(t) = \frac1N \frac1{\TK(\sigma)} \e^{-t/\TK(\sigma)} 
\tftrans(t) \biggbrak{1+2\sum_{q\geqs1}(-1)^q
\biggabs{\Gamma\biggpar{1+\frac{\pi\icx q}{\lambda T}}}^2 
\cos\biggpar{\frac{2\pi qt}T}},
\end{equation}
where $\tftrans(t)$ has the same behaviour as $\ftrans(t,0)$. The periodic
part of the distribution is minimal in multiples of $T$, and maximal in odd
multiples of $T/2$, while its nonperiodic part behaves like the non-periodic
part of the first-passage-time density. 

%%%%%%%%%%%%%%%%%%%%%%%%%%%%%%%%%%%%%%%%%%%%%%%%%%%%%%%%%%%%%%%%%%%%%%%%%%%%%%

\section{Conclusion}

The most important aspect of our rigorous expression for the residence-time
distribution is the fact that it is governed essentially by two
dimensionless parameters, $\lambda T$ and $T/\TK(\sigma)$, which can be
modified independently. The ratio $T/\TK(\sigma)$ between period and
Kramers time appears in most quantitative measures of SR, which indicate an
optimal amplification when $T$ is close to $2\TK(\sigma)$. In this regime,
the probability of transitions between potential wells becomes significant
during each period. The parameter $\lambda T$, by contrast, controls the
concentration of residence times within each period. Large values of
$\lambda T$ yield a sharply peaked residence-time distribution, regardless
of the peak's relative height. 

%%%%%%%%%%%%%%%%%%%%%%%%%%%%%%%%%%%%%%%%%%%%%%%%%%%%%%%%%%%%%%%%%%%%%%%%%%%%%%

\acknowledgments

It's a pleasure to thank Anton Bovier, Arkady Pikovsky and Dan Stein for
valuable discussions. 
N.B. thanks the WIAS for kind hospitality and financial support. B.G.
thanks the Universit\'e du Sud Toulon--Var and the CPT--CNRS Luminy for
kind hospitality. Financial support by the ESF Programme {\it Phase
Transitions and Fluctuation Phenomena for Random Dynamics in Spatially
Extended Systems (RDSES)\/} is gratefully acknowledged.

%%%%%%%%%%%%%%%%%%%%%%%%%%%%%%%%%%%%%%%%%%%%%%%%%%%%%%%%%%%%%%%%%%%%%%%%%%%%%%

\bibliography{BIB}
\bibliographystyle{unsrt}

%%%%%%%%%%%%%%%%%%%%%%%%%%%%%%%%%%%%%%%%%%%%%%%%%%%%%%%%%%%%%%%%%%%%%%%%%%%%%%

\end{document}